\documentclass{article}

\usepackage[final,nonatbib]{neurips_2022_ml4ps}

\DeclareUnicodeCharacter{2009}{\,}

\usepackage{graphicx}
\usepackage[utf8]{inputenc} 
\usepackage[T1]{fontenc}    
\usepackage{hyperref}       
\usepackage{url}            
\usepackage{booktabs}       
\usepackage{amsfonts}       
\usepackage{nicefrac}       
\usepackage{microtype}      
\usepackage{xcolor}         
\usepackage{biblatex}
\title{Elements of effective machine learning datasets in astronomy}

\author{%
  Bernie Boscoe* \\
  Department of Computer Science\\
  Occidental College\\
  Los Angeles, CA 90041 \\
  \texttt{boscoe@oxy.edu} \\
   \And
   Tuan Do \\
   Department of Physics and Astronomy \\
   UCLA \\
   Los Angeles, CA 90025 \\
   \texttt{tdo@astro.ucla.edu} \\
   \AND
   Evan Jones \\
   Department of Physics and Astronomy \\
   UCLA \\
   Los Angeles, CA 90025 \\
   \texttt{evan.jones@astro.ucla.edu} \\
   \And
   Yunqi Li \\
   Department of Physics and Astronomy \\
   UCLA \\
   Los Angeles, CA 90025 \\
   \And
   Kevin Alfaro \\
   Department of Physics and Astronomy \\
   UCLA \\
   Los Angeles, CA 90025 \\
   \And
   Christy Ma \\
   Department of Physics and Astronomy \\
   UCLA \\
   Los Angeles, CA 90025 \\
}
\begin{document}
\maketitle
\begin{abstract}
In this work, we identify elements of effective machine learning datasets in astronomy and present suggestions for their design and creation. Machine learning has become an increasingly important tool for analyzing and understanding the large-scale flood of data in astronomy. To take advantage of these tools, datasets are required for training and testing. However, building machine learning datasets for astronomy can be challenging. Astronomical data is collected from instruments built to explore science questions in a traditional fashion rather than to conduct machine learning. Thus, it is often the case that raw data, or even downstream processed data is not in a form amenable to machine learning. We explore the construction of machine learning datasets and we ask: what elements define effective machine learning datasets? We define effective machine learning datasets in astronomy to be formed with well-defined data points, structure, and metadata. We discuss why these elements are important for astronomical applications and ways to put them in practice. We posit that these qualities not only make the data suitable for machine learning, they also help to foster usable, reusable, and replicable science practices.

\end{abstract}

\section{Introduction}

In recent years astronomy has seen a wide application of machine learning (ML) in numerous subfields, from exoplanets and stellar astrophysics to extragalactic and cosmology applications \cite{dieleman_rotation-invariant_2015} \cite{walmsley_galaxy_2021}. In particular, the imminent start of the largest sky surveys ever conducted have motivated astronomers to adopt machine learning methods to filter, analyze, and extract information from these surveys. A large number of astronomy publications cite the Legacy Survey of Space and Time (LSST) and the Euclid space mission as the main drivers for implementing machine learning processes \cite{ivezic_lsst_2018} \cite{racca_euclid_2016}. The sheer volume of data to be generated by those telescopes is far more than any other survey or mission to date \cite{acquaviva_pushing_2019} \cite{wu_predicting_2020} \cite{sanchez_lsst_2020} \cite{mandelbaum_weak_2018} \cite{newman_photometric_2022}. 

Astronomy as a field has famously and painstakingly made numerous datasets freely available in the forms of mission archives and survey repositories, but there has been less work in general on defining and building common practices for creating machine learning datasets \cite{gebru_datasheets_2018}. We claim traditional  mission and survey catalogs are effective astronomy datasets: they comprise careful calibrations, calculations, meticulously detailed descriptions and imagery, typically in a format enabling SQL queries \cite{szalay_designing_2000}. They are queryable, available, interpretable for humans and readable by astronomers’ tools. 
To employ machine learning techniques, these datasets are commonly transformed for tools developed in industry like PyTorch and TensorFlow \cite{tensorflow2015-whitepaper}\cite{NEURIPS2019_9015}. This laborious transformation process to an ML-ready dataset requires choices about what content, structure, and metadata to include. However, little work has been done on understanding this process and identifying what constitutes an effective machine learning dataset for astronomy \cite{peek_search_2021}. 

In this paper, we outline three elements in effective machine learning datasets for astronomy and suggest adopting them in practice. In our view, effective datasets are useful, usable, and reusable for all researchers.  We propose that effective ML datasets have three characteristics: well-defined data points, well-defined structure, and well-defined metadata. In the sections below, we explore these three characteristics through the lens of astronomy datasets for machine learning. We use the term well defined to describe datasets conceptually, structurally, and holistically. The datasets should possess the necessary characteristics for being effective, useful, usable, and reusable \cite{weijmans_streamlining_nodate}.

\begin{figure}[!h]
\centering
\includegraphics[width=5.5in]{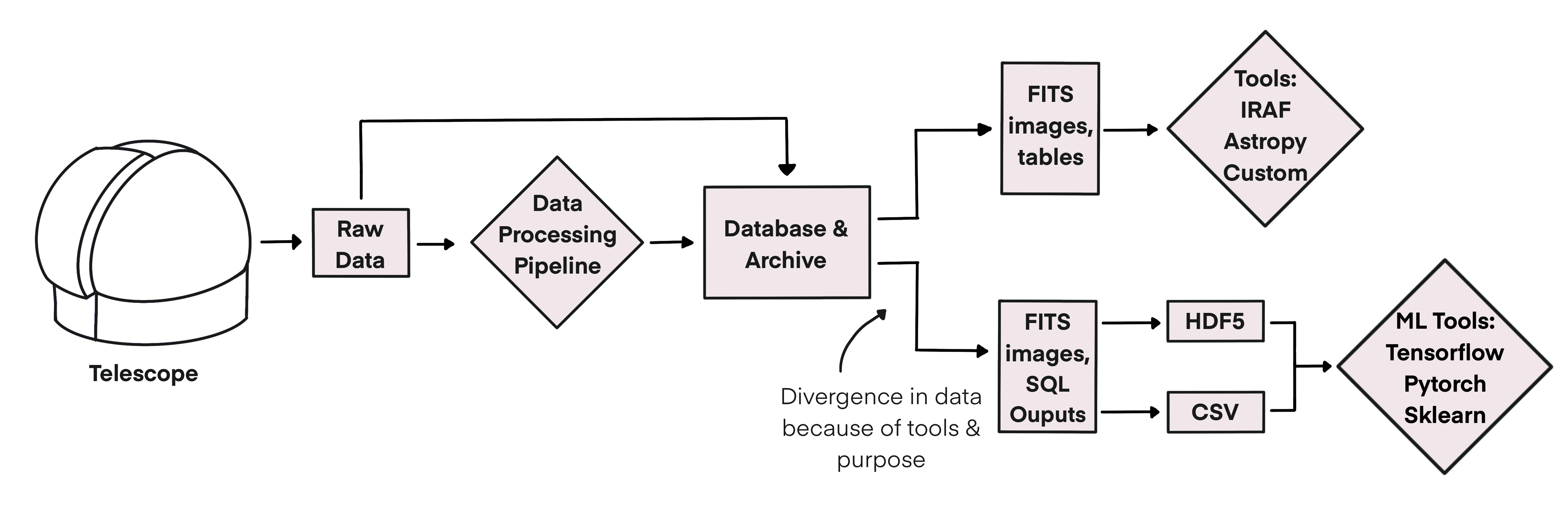}
\caption{\footnotesize 
Diagram of a typical data flow for astronomical data. From the left, observations from a telescope, resulting data products are fed into scientific analysis tools. Divergences in the data flow often occur when data intended for machine learning applications take one form and data used by more traditional astronomy tools take another. Divergence occurs because 1)  science goals affect the selection and quantity of data and, 2) data requirements for tools may differ. 
}
\label{fig:diagram}
\end{figure}

\section{Well-defined data points}

Astronomical data in an upstream state often consists of unstructured raw images and spectra from telescopes that require further processing into data products for use in scientific analysis. Additionally, further processing is required to transform these data products into structured data points (tabular, time series, etc.) or image formats for machine learning. Some areas of research do not use image data, while others use combinations of both image and measurement data. In this paper we provide best practices using both; but image data can be eliminated and the data flows would remain similar.

Well-defined data points are outcomes of processing raw data, and have clear boundaries delineating what information is and is not contained in the dataset. Some boundaries stem from choices made by the people  previously or presently involved in the dataset creation process, while other boundaries come from structural limitations such as instrument configurations and the size and construction of the mirrors. These boundaries may be mutable or immutable, depending on the context; our focus is on documenting these decision-making processes as an integral part of ML datasets, as associated metadata.

Structural boundaries describe 'fixed' data collected and stored from instruments, wavelengths, time durations, and regions of sky, to name a few. Boundaries resulting from human choice, such the selection of objects, are malleable; and can be reworked to create different datasets. Datasets are blended results of iterations of decisions. For example, a machine learning dataset might include galaxies over a particular region of the sky. The selection of galaxies is a choice while the region of the sky may be limited by the original survey parameters. Documenting details for why a particular dataset's attributes were chosen are helpful for others to understand and potentially reuse data. In this sense, the boundaries can define a limited ground truth. Well-defined data attributes formed from careful documentation and decision-making processes should be documented and consistent. This allows astronomers to have confidence in the use and reuse of the data and in understanding potential biases in the data. 

When selecting and organizing data points, astronomers must make decisions about:
1) quantifying the quality of data points, 2) establishing criteria for included data points, 3) establishing outlier criteria, and 4) identifying and potentially removing missing or low-quality data.

Data quality measures are a way to signify how much trust the data curators have on the resulting data points. For example, some data points have higher uncertainties or have an increased chance of being artifacts due to limits of the instruments used to make the measurements. By including attributes such as data quality flags \cite{aihara_second_2019}, data curators can help users filter data according to quality that might be necessary for different ML models. For example, some machine learning models are much more sensitive to noisy data than others. 

Outliers are important to identify and potentially remove. Two main types of outliers are: (1) data points outside of the typical sample distribution and (2) data points that are erroneous measurements. In machine learning, training data that contains outliers can strongly bias the predictions of models. Outliers that are confirmed to be artifacts or errors that make them ‘mistakes’ should likely be discarded, as machine learning algorithms will ‘learn’ these errors and skew results. A different approach must be taken with outliers that are likely to have been measured correctly but have values that are very far away from the rest of the data distribution. This type of outlier should be kept in the data but considerations must be taken about how to deal with it. Identification of these outliers require characterizing the statistical distribution of the existing data points. Documenting these outliers will alert users to their potential effects during ML training. When dealing with outliers, one approach users can take is to bin them as a category, thereby reducing their effect or otherwise diminishing their power with other statistical techniques.

Missing data points are also important to consider. Missing data might be in the form of a single attribute of a data point, or, it might mean a large sector of data points is not available. This might happen because of instrumentation limitations. For example, some astronomical datasets may contain both imaging and spectroscopic observations of many objects. However, because spectroscopy is much more difficult to obtain than images, some objects may be missing spectroscopic data. One way to mitigate problematic missing data is to use simulated data \cite{ciprijanovic_deepmerge_2020}. In machine learning, interpolation and extrapolation techniques can mitigate these issues, with varying degrees of success. Sampling techniques such as these may shape a better representation of a possible ground truth for a particular study. Astronomers should interpret results sensitive to missing data as a way of handling known limitations.

\section{Well-defined dataset structure  }

The form and structure of datasets are important for both machine learning and astronomy, but additional work must be taken to translate astronomical data points into ML-ready datasets. Structural considerations such as the data format, tabular shape, and image sizes and dimensions can have major impacts on the type of machine learning models able to be used. 

In astronomy, data is often stored in the FITS file format, which is a broad data format that can store metadata, imaging, spectra, and tabular data \cite{greisen_fits_2002}\cite{scroggins_once_2020}. While FITS files are highly compatible with, and designed for astronomical software, the files are not standardized enough to use directly in machine learning models.  Machine learning tools expect inputs of certain data types, and are not flexible enough to handle aberrations in data structures.

While there are API libraries available such as AstroPy \cite{collaboration_astropy_2022} to convert FITS files into machine learning amenable data formats, this transformation is not trivial, and can be problematic. PNG files are the most common image format used in mainstream machine learning, with lower resolution and smaller dimensional images comprising many datasets such as ImageNet or MNIST. PNG and JPG file formats use at the smallest 8 bits per channel, up to three channels (RGB). FITS image files, on the other hand, may contain many channels of various wavelengths at once, for example the COSMOS survey \cite{scoville_cosmic_2007}\cite{laigle_cosmos2015_2016} contains thirty different bands of wavelengths for its image data.

Reducing image resolution, number of channels, and image dimensions can greatly affect the amount of information contained in each image. For example, astronomers use multiple units for brightness, but in transformations to PNG these values would be reduced to a scale of integers from 0-255, greatly reducing each measurement's precision. Therefore, it is not recommended in many cases to transform astronomy images stored in FITS files to PNGs or JPGs. Also, FITS holds metadata about each image in its header section which also needs to be retained, and would be lost in a format transformation.

Instead of PNG files, a more flexible format that works well with machine learning algorithms like HDF5 \cite{noauthor_hdf5_2022} is a wiser choice. HDF5 provides a better compromise between preserving all the information in FITS and ease of use in ML models.
HDF5 is a library and associated file format that can store most types of multi-dimensional array data. This means it is able to preserve all wavelength channels in an astronomical image. The HDF5 data format is capable of storing metadata, tabular data, and image data, and works well to enable data to be ingested into TensorFlow or PyTorch. 
One major downside to HDF5 for astronomers is that its flexibility of storing metadata and data varies widely in terms of structure and labeling. Therefore, the FITS structure astronomers are familiar with is absent in HDF5, and conversions between the two formats are not one-to-one mappings.

Because of this issue, datasets for machine learning in astronomy must retain critical information spanning two file formats. Best practices include providing code to produce the HDF5 file in addition to descriptions of the storage of metadata alongside the data itself. There is no elegant solution; perhaps in the future a more seamless system could be developed to ingest FITS images into machine learning, but image sizes and the number of channels could be too large for model training. Issues of computing power, training time, and file size are of serious importance when constructing ML datasets.

Restrictions imposed on data formats originate from tool design outside of astronomy, so the structure of machine learning datasets must be made to align the science goals with requirements from the tools. For instance, computer vision for recognizing galaxy morphologies may need to work for galaxies of many different scale sizes, but ML tools require images to be a specific dimension. In most cases, the images from telescopes are much larger than the sizes that are accepted by ML tools. Structuring astronomical data for current ML tools can limit the amount of information that was originally available, for example  in mainstream ML tools, to decrease training time image dimensions and resolutions are routinely resized and downscaled. Astronomers building ML datasets should consider and articulate potential information loss from transforming the data.

\section{Well-defined metadata}
Well-defined metadata includes: 1) all contextual information relevant to the creation of the dataset, 2) the features and form of the dataset, 3) motivations for creating the dataset with respect to the initial scientific goal. 

To document contextual information, an astronomer should document the creation of the dataset starting from how the original data was obtained from the archives or instruments to the dataset’s final form. For example, users of most astronomy archives use SQL queries to extract subsets of data; queries used to form ML datasets should be preserved. Filtering and processing steps should also be documented and explained. Often, ML datasets in astronomy end up being different versions of similar data; specific versioning schemas should be enacted to ensure consistency in subsequent use of each dataset. 

With respect to feature and form preservation, well-defined tabular metadata should include metadata for each column and should include details such as units, descriptions, and how features were obtained. For images from FITS files, relevant data from the FITS headers should be preserved outside of FITS, if possible, and made available with the final machine learning dataset. The file format version (if any) and tools that can read the dataset should be documented in an archival file format.

Metadata detailing the motivation for the creation of the dataset enables users to better understand ways it contextually could be used. Often, ML datasets for astronomy are created with specific science goals in mind (e.g. photometric redshifts \cite{jones_photometric_2022}). Decisions about the boundaries of the initial dataset are easier to understand if the original goals are known. By documenting the goals and the requirements of the initial scientific investigations, users can situate the dataset with respect to their own science goals. 

\section{Conclusion}

Datasets with well-defined data points, structure, and metadata invite reusability, which makes scientific studies more reliable and efficient. Because of the intensive labor that necessarily goes into creating machine learning-ready datasets, reusing these datasets makes enormous sense to researchers. Machine learning datasets possessing the elements we outlined in this work are a way forward to effectively use massive datasets that are unable to be manually explored. We recommend that astronomers create effective datasets with an eye toward machine learning-ready characteristics, for the benefit of all data-intensive science. 
\section{Broader Impacts}
We believe that our paper may have a positive impact in the field of astronomy, because we suggest best practices that can be incorporated into workflows to improve research practices surrounding the creation, curation, and preservation of datasets for machine learning in astronomy. We feel that potential ethical issues stemming from our work are minimal. 
\printbibliography

@incollection{greisen_fits_2002,
	series = {Astrophysics and {Space} {Science} {Library}},
	title = {{FITS}: {A} {Remarkable} {Achievement} in {Information} {Exchange}},
	copyright = {©2002 Kluwer Academic Publishers},
	isbn = {978-1-4020-1178-8},
	shorttitle = {{FITS}},
	url = {http://link.springer.com/chapter/10.1007/0-306-48080-8_5},
	language = {en},
	number = {285},

	booktitle = {Information {Handling} in {Astronomy} - {Historical} {Vistas}},
	publisher = {Springer Netherlands},
	author = {Greisen, E. W.},
	editor = {Heck, André},
	month = jan,
	year = {2002},
	note = {00002},
	pages = {71--87},
}

@article{wu_predicting_2020,
	title = {Predicting galaxy spectra from images with hybrid convolutional neural networks},
	url = {http://arxiv.org/abs/2009.12318},
	
	journal = {arXiv:2009.12318 [astro-ph]},
	author = {Wu, John F. and Peek, J. E. G.},
	month = nov,
	year = {2020},
	note = {arXiv: 2009.12318},
}

@article{walmsley_galaxy_2021,
	title = {Galaxy {Zoo} {DECaLS}: {Detailed} visual morphology measurements from volunteers and deep learning for 314 000 galaxies},
	volume = {509},
	issn = {0035-8711},
	shorttitle = {Galaxy {Zoo} {DECaLS}},
	url = {https://doi.org/10.1093/mnras/stab2093},
	doi = {10.1093/mnras/stab2093},
	number = {3},
	
	journal = {Monthly Notices of the Royal Astronomical Society},
	author = {Walmsley, Mike and Lintott, Chris and Géron, Tobias and Kruk, Sandor and Krawczyk, Coleman and Willett, Kyle W and Bamford, Steven and Kelvin, Lee S and Fortson, Lucy and Gal, Yarin and Keel, William and Masters, Karen L and Mehta, Vihang and Simmons, Brooke D and Smethurst, Rebecca and Smith, Lewis and Baeten, Elisabeth M and Macmillan, Christine},
	month = jan,
	year = {2021},
	pages = {3966--3988},
}

@article{laigle_cosmos2015_2016,
	title = {{THE} {COSMOS2015} {CATALOG}: {EXPLORING} {THE} 1 \&lt\${\textbackslash}mathsemicolon\$ \${\textbackslash}less\$i\${\textbackslash}greater\$z\${\textbackslash}less\$/i\${\textbackslash}greater\$ \&lt\${\textbackslash}mathsemicolon\$ 6 {UNIVERSE} {WITH} {HALF} {A} {MILLION} {GALAXIES}},
	volume = {224},
	issn = {0067-0049},
	shorttitle = {{THE} {COSMOS2015} {CATALOG}},
	url = {https://doi.org/10.3847/0067-0049/224/2/24},
	doi = {10.3847/0067-0049/224/2/24},
	language = {en},
	number = {2},
	
	journal = {The Astrophysical Journal Supplement Series},
	author = {Laigle, C. and McCracken, H. J. and Ilbert, O. and Hsieh, B. C. and Davidzon, I. and Capak, P. and Hasinger, G. and Silverman, J. D. and Pichon, C. and Coupon, J. and Aussel, H. and Borgne, D. Le and Caputi, K. and Cassata, P. and Chang, Y.-Y. and Civano, F. and Dunlop, J. and Fynbo, J. and Kartaltepe, J. S. and Koekemoer, A. and Fèvre, O. Le and Floc'h, E. Le and Leauthaud, A. and Lilly, S. and Lin, L. and Marchesi, S. and Milvang-Jensen, B. and Salvato, M. and Sanders, D. B. and Scoville, N. and Smolcic, V. and Stockmann, M. and Taniguchi, Y. and Tasca, L. and Toft, S. and Vaccari, Mattia and Zabl, J.},
	month = jun,
	year = {2016},
	note = {Publisher: American Astronomical Society},
	pages = {24},
}

@article{dieleman_rotation-invariant_2015,
	title = {Rotation-invariant convolutional neural networks for galaxy morphology prediction},
	volume = {450},
	issn = {0035-8711},
	url = {https://doi.org/10.1093/mnras/stv632},
	doi = {10.1093/mnras/stv632},
	number = {2},

	journal = {Monthly Notices of the Royal Astronomical Society},
	author = {Dieleman, Sander and Willett, Kyle W. and Dambre, Joni},
	month = jun,
	year = {2015},
	pages = {1441--1459},
}

@article{aihara_second_2019,
	title = {Second {Data} {Release} of the {Hyper} {Suprime}-{Cam} {Subaru} {Strategic} {Program}},
	volume = {71},
	issn = {0004-6264, 2053-051X},
	url = {http://arxiv.org/abs/1905.12221},
	doi = {10.1093/pasj/psz103},
	number = {6},

	journal = {Publications of the Astronomical Society of Japan},
	author = {Aihara, Hiroaki and AlSayyad, Yusra and Ando, Makoto and Armstrong, Robert and Bosch, James and Egami, Eiichi and Furusawa, Hisanori and Furusawa, Junko and Goulding, Andy and Harikane, Yuichi and Hikage, Chiaki and Ho, Paul T. P. and Hsieh, Bau-Ching and Huang, Song and Ikeda, Hiroyuki and Imanishi, Masatoshi and Ito, Kei and Iwata, Ikuru and Jaelani, Anton T. and Kakuma, Ryota and Kawana, Kojiro and Kikuta, Satoshi and Kobayashi, Umi and Koike, Michitaro and Komiyama, Yutaka and Li, Xiangchong and Liang, Yongming and Lin, Yen-Ting and Luo, Wentao and Lupton, Robert and Lust, Nate B. and MacArthur, Lauren A. and Matsuoka, Yoshiki and Mineo, Sogo and Miyatake, Hironao and Miyazaki, Satoshi and More, Surhud and Murata, Ryoma and Namiki, Shigeru V. and Nishizawa, Atsushi J. and Oguri, Masamune and Okabe, Nobuhiro and Okamoto, Sakurako and Okura, Yuki and Ono, Yoshiaki and Onodera, Masato and Onoue, Masafusa and Osato, Ken and Ouchi, Masami and Shibuya, Takatoshi and Strauss, Michael A. and Sugiyama, Naoshi and Suto, Yasushi and Takada, Masahiro and Takagi, Yuhei and Takata, Tadafumi and Takita, Satoshi and Tanaka, Masayuki and Terai, Tsuyoshi and Toba, Yoshiki and Uchiyama, Hisakazu and Utsumi, Yousuke and Wang, Shiang-Yu and Wang, Wenting and Yamada, Yoshihiko},
	month = dec,
	year = {2019},
	note = {arXiv:1905.12221 [astro-ph]},
	pages = {114},
}

@misc{ivezic_lsst_2018,
	title = {{LSST}: from {Science} {Drivers} to {Reference} {Design} and {Anticipated} {Data} {Products}},
	shorttitle = {{LSST}},
	url = {http://arxiv.org/abs/0805.2366},
	doi = {10.3847/1538-4357/ab042c},
	
	author = {Ivezić, Željko and Kahn, Steven M. and Tyson, J. Anthony and Abel, Bob and Acosta, Emily and Allsman, Robyn and Alonso, David and AlSayyad, Yusra and Anderson, Scott F. and Andrew, John and Angel, James Roger P. and Angeli, George Z. and Ansari, Reza and Antilogus, Pierre and Araujo, Constanza and Armstrong, Robert and Arndt, Kirk T. and Astier, Pierre and Aubourg, Éric and Auza, Nicole and Axelrod, Tim S. and Bard, Deborah J. and Barr, Jeff D. and Barrau, Aurelian and Bartlett, James G. and Bauer, Amanda E. and Bauman, Brian J. and Baumont, Sylvain and Becker, Andrew C. and Becla, Jacek and Beldica, Cristina and Bellavia, Steve and Bianco, Federica B. and Biswas, Rahul and Blanc, Guillaume and Blazek, Jonathan and Blandford, Roger D. and Bloom, Josh S. and Bogart, Joanne and Bond, Tim W. and Borgland, Anders W. and Borne, Kirk and Bosch, James F. and Boutigny, Dominique and Brackett, Craig A. and Bradshaw, Andrew and Brandt, William Nielsen and Brown, Michael E. and Bullock, James S. and Burchat, Patricia and Burke, David L. and Cagnoli, Gianpietro and Calabrese, Daniel and Callahan, Shawn and Callen, Alice L. and Chandrasekharan, Srinivasan and Charles-Emerson, Glenaver and Chesley, Steve and Cheu, Elliott C. and Chiang, Hsin-Fang and Chiang, James and Chirino, Carol and Chow, Derek and Ciardi, David R. and Claver, Charles F. and Cohen-Tanugi, Johann and Cockrum, Joseph J. and Coles, Rebecca and Connolly, Andrew J. and Cook, Kem H. and Cooray, Asantha and Covey, Kevin R. and Cribbs, Chris and Cui, Wei and Cutri, Roc and Daly, Philip N. and Daniel, Scott F. and Daruich, Felipe and Daubard, Guillaume and Daues, Greg and Dawson, William and Delgado, Francisco and Dellapenna, Alfred and de Peyster, Robert and de Val-Borro, Miguel and Digel, Seth W. and Doherty, Peter and Dubois, Richard and Dubois-Felsmann, Gregory P. and Durech, Josef and Economou, Frossie and Eracleous, Michael and Ferguson, Henry and Figueroa, Enrique and Fisher-Levine, Merlin and Focke, Warren and Foss, Michael D. and Frank, James and Freemon, Michael D. and Gangler, Emmanuel and Gawiser, Eric and Geary, John C. and Gee, Perry and Geha, Marla and Gessner, Charles J. B. and Gibson, Robert R. and Gilmore, D. Kirk and Glanzman, Thomas and Glick, William and Goldina, Tatiana and Goldstein, Daniel A. and Goodenow, Iain and Graham, Melissa L. and Gressler, William J. and Gris, Philippe and Guy, Leanne P. and Guyonnet, Augustin and Haller, Gunther and Harris, Ron and Hascall, Patrick A. and Haupt, Justine and Hernandez, Fabio and Herrmann, Sven and Hileman, Edward and Hoblitt, Joshua and Hodgson, John A. and Hogan, Craig and Huang, Dajun and Huffer, Michael E. and Ingraham, Patrick and Innes, Walter R. and Jacoby, Suzanne H. and Jain, Bhuvnesh and Jammes, Fabrice and Jee, James and Jenness, Tim and Jernigan, Garrett and Jevremović, Darko and Johns, Kenneth and Johnson, Anthony S. and Johnson, Margaret W. G. and Jones, R. Lynne and Juramy-Gilles, Claire and Jurić, Mario and Kalirai, Jason S. and Kallivayalil, Nitya J. and Kalmbach, Bryce and Kantor, Jeffrey P. and Karst, Pierre and Kasliwal, Mansi M. and Kelly, Heather and Kessler, Richard and Kinnison, Veronica and Kirkby, David and Knox, Lloyd and Kotov, Ivan V. and Krabbendam, Victor L. and Krughoff, K. Simon and Kubánek, Petr and Kuczewski, John and Kulkarni, Shri and Ku, John and Kurita, Nadine R. and Lage, Craig S. and Lambert, Ron and Lange, Travis and Langton, J. Brian and Guillou, Laurent Le and Levine, Deborah and Liang, Ming and Lim, Kian-Tat and Lintott, Chris J. and Long, Kevin E. and Lopez, Margaux and Lotz, Paul J. and Lupton, Robert H. and Lust, Nate B. and MacArthur, Lauren A. and Mahabal, Ashish and Mandelbaum, Rachel and Marsh, Darren S. and Marshall, Philip J. and Marshall, Stuart and May, Morgan and McKercher, Robert and McQueen, Michelle and Meyers, Joshua and Migliore, Myriam and Miller, Michelle and Mills, David J. and Miraval, Connor and Moeyens, Joachim and Monet, David G. and Moniez, Marc and Monkewitz, Serge and Montgomery, Christopher and Mueller, Fritz and Muller, Gary P. and Arancibia, Freddy Muñoz and Neill, Douglas R. and Newbry, Scott P. and Nief, Jean-Yves and Nomerotski, Andrei and Nordby, Martin and O'Connor, Paul and Oliver, John and Olivier, Scot S. and Olsen, Knut and O'Mullane, William and Ortiz, Sandra and Osier, Shawn and Owen, Russell E. and Pain, Reynald and Palecek, Paul E. and Parejko, John K. and Parsons, James B. and Pease, Nathan M. and Peterson, J. Matt and Peterson, John R. and Petravick, Donald L. and Petrick, M. E. Libby and Petry, Cathy E. and Pierfederici, Francesco and Pietrowicz, Stephen and Pike, Rob and Pinto, Philip A. and Plante, Raymond and Plate, Stephen and Price, Paul A. and Prouza, Michael and Radeka, Veljko and Rajagopal, Jayadev and Rasmussen, Andrew P. and Regnault, Nicolas and Reil, Kevin A. and Reiss, David J. and Reuter, Michael A. and Ridgway, Stephen T. and Riot, Vincent J. and Ritz, Steve and Robinson, Sean and Roby, William and Roodman, Aaron and Rosing, Wayne and Roucelle, Cecille and Rumore, Matthew R. and Russo, Stefano and Saha, Abhijit and Sassolas, Benoit and Schalk, Terry L. and Schellart, Pim and Schindler, Rafe H. and Schmidt, Samuel and Schneider, Donald P. and Schneider, Michael D. and Schoening, William and Schumacher, German and Schwamb, Megan E. and Sebag, Jacques and Selvy, Brian and Sembroski, Glenn H. and Seppala, Lynn G. and Serio, Andrew and Serrano, Eduardo and Shaw, Richard A. and Shipsey, Ian and Sick, Jonathan and Silvestri, Nicole and Slater, Colin T. and Smith, J. Allyn and Smith, R. Chris and Sobhani, Shahram and Soldahl, Christine and Storrie-Lombardi, Lisa and Stover, Edward and Strauss, Michael A. and Street, Rachel A. and Stubbs, Christopher W. and Sullivan, Ian S. and Sweeney, Donald and Swinbank, John D. and Szalay, Alexander and Takacs, Peter and Tether, Stephen A. and Thaler, Jon J. and Thayer, John Gregg and Thomas, Sandrine and Thukral, Vaikunth and Tice, Jeffrey and Trilling, David E. and Turri, Max and Van Berg, Richard and Berk, Daniel Vanden and Vetter, Kurt and Virieux, Francoise and Vucina, Tomislav and Wahl, William and Walkowicz, Lucianne and Walsh, Brian and Walter, Christopher W. and Wang, Daniel L. and Wang, Shin-Yawn and Warner, Michael and Wiecha, Oliver and Willman, Beth and Winters, Scott E. and Wittman, David and Wolff, Sidney C. and Wood-Vasey, W. Michael and Wu, Xiuqin and Xin, Bo and Yoachim, Peter and Zhan, Hu},
	month = may,
	year = {2018},
	note = {arXiv:0805.2366 [astro-ph]},
}

@inproceedings{racca_euclid_2016,
	title = {The {Euclid} mission design},
	url = {http://arxiv.org/abs/1610.05508},
	doi = {10.1117/12.2230762},
	
	author = {Racca, Giuseppe D. and Laureijs, Rene and Stagnaro, Luca and Salvignol, Jean Christophe and Alvarez, Jose Lorenzo and Criado, Gonzalo Saavedra and Venancio, Luis Gaspar and Short, Alex and Strada, Paolo and Boenke, Tobias and Colombo, Cyril and Calvi, Adriano and Maiorano, Elena and Piersanti, Osvaldo and Prezelus, Sylvain and Rosato, Pierluigi and Pinel, Jacques and Rozemeijer, Hans and Lesna, Valentina and Musi, Paolo and Sias, Marco and Anselmi, Alberto and Cazaubiel, Vincent and Vaillon, Ludovic and Mellier, Yannick and Amiaux, Jerome and Berthe, Michel and Sauvage, Marc and Azzollini, Ruyman and Cropper, Mark and Pottinger, Sabrina and Jahnke, Knud and Ealet, Anne and Maciaszek, Thierry and Pasian, Fabio and Zacchei, Andrea and Scaramella, Roberto and Hoar, John and Kohley, Ralf and Vavrek, Roland and Rudolph, Andreas and Schmidt, Micha},
	month = jul,
	year = {2016},
	note = {arXiv:1610.05508 [astro-ph]},
	pages = {99040O},
}

@article{sanchez_lsst_2020,
	title = {The {LSST} {DESC} data challenge 1: generation and analysis of synthetic images for next-generation surveys},
	volume = {497},
	issn = {0035-8711},
	shorttitle = {The {LSST} {DESC} data challenge 1},
	url = {https://doi.org/10.1093/mnras/staa1957},
	doi = {10.1093/mnras/staa1957},
	number = {1},
	
	journal = {Monthly Notices of the Royal Astronomical Society},
	author = {Sánchez, J and Walter, C W and Awan, H and Chiang, J and Daniel, S F and Gawiser, E and Glanzman, T and Kirkby, D and Mandelbaum, R and Slosar, A and Wood-Vasey, W M and AlSayyad, Y and Burke, C J and Digel, S W and Jarvis, M and Johnson, T and Kelly, H and Krughoff, S and Lupton, R H and Marshall, P J and Peterson, J R and Price, P A and Sembroski, G and Van Klaveren, B and Wiesner, M P and Xin, B and {The LSST Dark Energy Science Collaboration}},
	month = sep,
	year = {2020},
	pages = {210--228},
}

@article{mandelbaum_weak_2018,
	title = {Weak {Lensing} for {Precision} {Cosmology}},
	volume = {56},
	url = {https://doi.org/10.1146/annurev-astro-081817-051928},
	doi = {10.1146/annurev-astro-081817-051928},
	number = {1},

	journal = {Annual Review of Astronomy and Astrophysics},
	author = {Mandelbaum, Rachel},
	year = {2018},
	note = {\_eprint: https://doi.org/10.1146/annurev-astro-081817-051928},
	pages = {393--433},
}

@article{newman_photometric_2022,
	title = {Photometric {Redshifts} for {Next}-{Generation} {Surveys}},
	volume = {60},
	url = {https://doi.org/10.1146/annurev-astro-032122-014611},
	doi = {10.1146/annurev-astro-032122-014611},
	number = {1},

	journal = {Annual Review of Astronomy and Astrophysics},
	author = {Newman, Jeffrey A. and Gruen, Daniel},
	year = {2022},
	note = {\_eprint: https://doi.org/10.1146/annurev-astro-032122-014611},
	pages = {363--414},
}

@inproceedings{szalay_designing_2000,
	title = {Designing and mining multi-terabyte astronomy archives: the {Sloan} {Digital} {Sky} {Survey}},
	isbn = {978-1-58113-217-5},
	shorttitle = {Designing and mining multi-terabyte astronomy archives},
	url = {http://portal.acm.org/citation.cfm?doid=342009.335439},
	doi = {10.1145/342009.335439},
	language = {en},

	publisher = {ACM Press},
	author = {Szalay, Alexander S. and Kunszt, Peter Z. and Thakar, Ani and Gray, Jim and Slutz, Don and Brunner, Robert J.},
	year = {2000},
	pages = {451--462},
}

@article{acquaviva_pushing_2019,
	title = {Pushing the {Technical} {Frontier}: {From} {Overwhelmingly} {Large} {Data} {Sets} to {Machine} {Learning}},
	volume = {15},
	issn = {1743-9213, 1743-9221},
	shorttitle = {Pushing the {Technical} {Frontier}},
	url = {http://arxiv.org/abs/1901.05978},
	doi = {10.1017/S1743921319003077},
	number = {S341},
	
	journal = {Proceedings of the International Astronomical Union},
	author = {Acquaviva, Viviana},
	month = nov,
	year = {2019},
	note = {arXiv:1901.05978 [astro-ph]},
	pages = {88--98},
}

@article{peek_search_2021,
	title = {Search {By} {Image}: {Citizen} {Science} and {Deep} {Learning} for next-generation archives},
	volume = {53},
	shorttitle = {Search {By} {Image}},
	url = {https://ui.adsabs.harvard.edu/abs/2021AAS...23830106P},
	
	author = {Peek, J. and White, R.},
	month = jun,
	year = {2021},
	note = {Conference Name: American Astronomical Society Meeting Abstracts
ADS Bibcode: 2021AAS...23830106P},
	pages = {301.06},
}

@article{ciprijanovic_deepmerge_2020,
	title = {{DeepMerge}: {Classifying} high-redshift merging galaxies with deep neural networks},
	volume = {32},
	issn = {2213-1337},
	shorttitle = {{DeepMerge}},
	
	doi = {10.1016/j.ascom.2020.100390},
	language = {en},

	journal = {Astronomy and Computing, Volume 32, article id. 100390.},
	author = {Ćiprijanović, A. and Snyder, G. F. and Nord, B. and Peek, J. E. G.},
	month = jul,
	year = {2020},
	pages = {100390},
}

@article{weijmans_streamlining_nodate,
	title = {Streamlining the {Sloan} {Digital} {Sky} {Survey} {Public} {Data} {Releases}: {Changes} {Made} and {Lessons} {Learned}},
	language = {en},
	author = {Weijmans, Anne-Marie and Brownstein, Joel and Thakar, Ani and Blanton, Michael and Cherinka, Brian and Masters, Karen and Raddick, M Jordan},
	pages = {4},
}

@article{gebru_datasheets_2018,
	title = {Datasheets for {Datasets}},
	url = {http://arxiv.org/abs/1803.09010},

	journal = {arXiv:1803.09010 [cs]},
	author = {Gebru, Timnit and Morgenstern, Jamie and Vecchione, Briana and Vaughan, Jennifer Wortman and Wallach, Hanna and Daumeé III, Hal and Crawford, Kate},
	month = mar,
	year = {2018},
	note = {arXiv: 1803.09010},
}

@misc{tensorflow2015-whitepaper,
title={ {TensorFlow}: Large-Scale Machine Learning on Heterogeneous Systems},
url={https://www.tensorflow.org/},
note={Software available from tensorflow.org},
author={
    Mart\'{i}n~Abadi and
    Ashish~Agarwal and
    Paul~Barham and
    Eugene~Brevdo and
    Zhifeng~Chen and
    Craig~Citro and
    Greg~S.~Corrado and
    Andy~Davis and
    Jeffrey~Dean and
    Matthieu~Devin and
    Sanjay~Ghemawat and
    Ian~Goodfellow and
    Andrew~Harp and
    Geoffrey~Irving and
    Michael~Isard and
    Yangqing Jia and
    Rafal~Jozefowicz and
    Lukasz~Kaiser and
    Manjunath~Kudlur and
    Josh~Levenberg and
    Dandelion~Man\'{e} and
    Rajat~Monga and
    Sherry~Moore and
    Derek~Murray and
    Chris~Olah and
    Mike~Schuster and
    Jonathon~Shlens and
    Benoit~Steiner and
    Ilya~Sutskever and
    Kunal~Talwar and
    Paul~Tucker and
    Vincent~Vanhoucke and
    Vijay~Vasudevan and
    Fernanda~Vi\'{e}gas and
    Oriol~Vinyals and
    Pete~Warden and
    Martin~Wattenberg and
    Martin~Wicke and
    Yuan~Yu and
    Xiaoqiang~Zheng},
  year={2015},
}
\end{document}